\documentclass[sigconf, nonacm]{acmart}

\newcommand\vldbdoi{XX.XX/XXX.XX}
\newcommand\vldbpages{XXX-XXX}
\newcommand\vldbvolume{14}
\newcommand\vldbissue{1}
\newcommand\vldbyear{2020}
\newcommand\vldbauthors{\authors}
\newcommand\vldbtitle{\shorttitle} 
\newcommand\vldbavailabilityurl{URL_TO_YOUR_ARTIFACTS}
\newcommand\vldbpagestyle{plain} 

\usepackage{pythonhighlight}
\usepackage{booktabs}
\usepackage{xcolor}
\usepackage{fancyvrb}
\usepackage{listings}
\usepackage{multirow}
\usepackage{xspace}
\usepackage{mdframed}
\usepackage{csquotes}
\usepackage{subcaption}
\usepackage{graphicx}
\usepackage{tabularray}
\usepackage{makecell}

\definecolor{darkgreen}{rgb}{0,0.5,0}
\definecolor{purple}{rgb}{0.75,0,0.75}
\definecolor{brown}{rgb}{0.65,0.16,0.16}
\definecolor{darkslateblue}{rgb}{0.28, 0.24, 0.55}
\definecolor{orange}{rgb}{1.0, 0.647, 0}
\definecolor{softpink}{rgb}{0.906, 0.329, 0.502}
\definecolor{neonblue}{rgb}{0.12, 0.32, 1.0}
\definecolor{unitedred}{rgb}{0.85, 0.16, 0.11}
\definecolor{ao(english)}{rgb}{0.0, 0.5, 0.0}
\definecolor{bistre}{rgb}{0.24, 0.17, 0.12}

\definecolor{color1}{HTML}{caa8f5}
\definecolor{color2}{HTML}{e7ecef}
\definecolor{color3}{HTML}{edf6f9}
\definecolor{color4}{HTML}{e8e4db}

\newcount\notenum
\newcommand{\papercomment}[3]        {\global\advance\notenum by 1\textsl{\bf\color{#2}$\blacksquare$[#1\the\notenum: #3]}}

\newcommand{\eg}{\textit{e.~g.},\xspace}

\newcommand{\logql}{LogQL\xspace}

\newcommand{\sysName}{\textsc{LogQL-LM}} %

\newcommand{\cemph}[1]{\noindent\begin{mdframed}[hidealllines=true,backgroundcolor=blue!5,  
innertopmargin=3pt,
innerbottommargin=3pt,
innerrightmargin=2pt,
innerleftmargin=2pt,
leftmargin = 0pt,
rightmargin = 0pt,
skipabove = -12pt,
skipbelow = -12pt
]{#1}%
\end{mdframed}\vspace{-0.5em}}

\AtBeginDocument{%
  \providecommand\BibTeX{{%
    \normalfont B\kern-0.5em{\scshape i\kern-0.25em b}\kern-0.8em\TeX}}}

\setcopyright{acmcopyright}
\copyrightyear{2018}
\acmYear{2018}
\acmDOI{XXXXXXX.XXXXXXX}

\acmConference[Conference acronym 'XX]{Make sure to enter the correct
  conference title from your rights confirmation emai}{June 03--05,
  2018}{Woodstock, NY}
\acmPrice{15.00}
\acmISBN{978-1-4503-XXXX-X/18/06}

\begin{document}

\title{\textit{Chatting with Logs}: An exploratory study\\ on Finetuning LLMs for LogQL}

\author{Vishwanath Seshagiri}
\affiliation{%
  \institution{Emory University }
  \city{Atlanta}
  \country{USA}
}

\author{Siddharth Balyan}
\affiliation{%
  \institution{Composio}
  \city{Bangalore}
  \country{India}
}

\author{Vaastav Anand}
\affiliation{%
  \institution{MPI SWS}
  \city{Saarbrucken}
  \country{Germany}
}

\author{Kaustubh Dhole}
\affiliation{%
  \institution{Emory University }
  \city{Atlanta}
  \country{USA}
}

\author{Ishan Sharma}
\affiliation{%
  \institution{San Jose State University}
  \city{San Jose}
  \country{USA}
}

\author{Avani Wildani}
\affiliation{%
  \institution{Cloudflare and Emory University}
  \city{Atlanta}
  \country{USA}
}

\author{José Cambronero}
\affiliation{%
  \institution{Google}
  \city{Atlanta}
  \country{USA}
}

\author{Andreas Züfle}
\affiliation{%
  \institution{Emory University}
  \city{Atlanta}
  \country{USA}
}

\renewcommand{\shortauthors}{}

\begin{abstract}

Logging is a critical function in modern distributed applications, but the lack of standardization in log query languages and formats creates significant challenges.  Developers currently must write ad hoc queries in platform-specific languages, requiring expertise in both the query language and application-specific log details -- an impractical expectation given the variety of platforms and volume of logs and applications.  While generating these queries with large language models (LLMs) seems intuitive, we show that current LLMs struggle with log-specific query generation due to the lack of exposure to domain-specific knowledge.

We propose a novel natural language (NL) interface to address these inconsistencies and aide log query generation, enabling developers to create queries in a target log query language by providing NL inputs.  We further introduce ~\textbf{NL2QL}, a manually annotated, real-world dataset of natural language questions paired with corresponding \logql queries spread across three log formats, to promote the training and evaluation of NL-to-loq query systems. 
Using NL2QL, we subsequently fine-tune and evaluate several state of the art LLMs, and demonstrate their improved capability to generate accurate \logql queries. We perform further ablation studies to demonstrate the effect of additional training data, and the transferability across different log formats.
In our experiments, we find up to 75\% improvement of finetuned models to generate \logql queries compared to non finetuned models.
\vspace{0.6cm}

\end{abstract}

\maketitle

\pagestyle{\vldbpagestyle}
\begingroup\small\noindent\raggedright\textbf{PVLDB Reference Format:}\\
\vldbauthors. \vldbtitle. PVLDB, \vldbvolume(\vldbissue): \vldbpages, \vldbyear.\\
\href{https://doi.org/\vldbdoi}{doi:\vldbdoi}
\endgroup
\begingroup
\renewcommand\thefootnote{}\footnote{\noindent
This work is licensed under the Creative Commons BY-NC-ND 4.0 International License. Visit \url{https://creativecommons.org/licenses/by-nc-nd/4.0/} to view a copy of this license. For any use beyond those covered by this license, obtain permission by emailing \href{mailto:info@vldb.org}{info@vldb.org}. Copyright is held by the owner/author(s). Publication rights licensed to the VLDB Endowment. \\
\raggedright Proceedings of the VLDB Endowment, Vol. \vldbvolume, No. \vldbissue\ %
ISSN 2150-8097. \\
\href{https://doi.org/\vldbdoi}{doi:\vldbdoi} \\
}\addtocounter{footnote}{-1}\endgroup

\ifdefempty{\vldbavailabilityurl}{}{
\vspace{.3cm}
\begingroup\small\noindent\raggedright\textbf{PVLDB Artifact Availability:}\\
The source code is available at \url{https://github.com/nl2logql/LogQLLM}, the dataset and models are hosted on Huggingface at \url{https://huggingface.co/nl-to-logql}, and the demo application can be accessed on this link: \url{https://llm-response-simulator-alt-glitch.replit.app/}
\endgroup
}

\section{Introduction}
\label{sec:introduction_new}
As modern 
web and mobile applications are increasingly deployed as microservices, observability data (\eg Metrics, Logs, Events, Traces, etc.) are collected from various applications during the application runtime, and the tooling used to collect and store this data increasingly plays a critical role in modern cloud deployments of systems~\cite{dynatracereport}. 
Observability tools are crucial for understanding how the systems work at scale as they provide insight into key tasks including predicting resource requirements \cite{multilevel2018picoreti, firm2020qiu, retro2015mace}, diagnosing faults \cite{sage2021gan, monitorrank2013kim, microscope2018lin, faultlocal2019zhou}, identifying security breaches \cite{security2021avila}, and performing regular system checks \cite{mismatches2022seshagiri}. 
Despite the importance of observability tooling, there is a lack of 
standardization of how a user interacts with the different tools~\cite{cncfdsls}.

Log data is typically collected and accessed through various proprietary platforms, each of which have their own query language \cite{cncfdsls}.  
Querying this data is invariably \textit{ad hoc} and challenging, involving exact matches to a log format, and keywords 
to search over a large database \cite{odms2021Karumuri, canopy2017kaldor, schemafirst2022shkuro}. 
Moreover, there is little to no syntactical overlap between these languages, necessitating significant developer retraining when moving between products.

\begin{figure}[t]
\centering
\begin{subfigure}{\linewidth}
\includegraphics[width=\columnwidth]{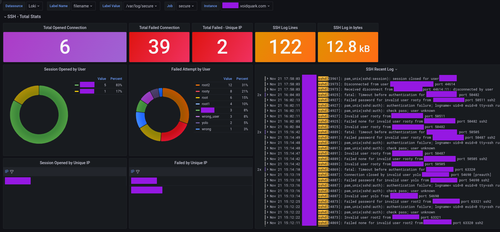}
\caption{Grafana Dashboard for OpenSSH logs retrieved from \cite{voidquark2024}\vspace{0.2cm}}    
\label{fig:grafanadashboard:image}
\end{subfigure}

\begin{subfigure}{\linewidth}
\begin{mdframed}[   
linecolor=blue!70!black,
backgroundcolor=color3,
linewidth=0pt,
roundcorner=10pt,
innerleftmargin=10pt,
innerrightmargin=10pt,
innertopmargin=6pt,
innerbottommargin=6pt,]

\begin{lstlisting}[
language=SQL,
basicstyle=\ttfamily\small\color{darkgreen},
]
sum by(instance) (count_over_time({
$label_name=~"$label_value", job=~"$job", 
instance=~"$instance"} |="sshd[" |=": session opened for"
| __error__="" [$__interval]))
\end{lstlisting}
\end{mdframed}
\caption{LogQL Query for calculating the total number of open connections (First panel)}
\label{fig:grafanadashboard:logqlquery}
\end{subfigure}

\caption{Example Grafana Dashboard}
\label{fig:grafanadashboard}
\end{figure}

Logs are often used to power insights dashboards.
For example, consider a Grafana dashboard for the OpenSSH application shown in \autoref{fig:grafanadashboard}. 
This dashboard comprises multiple panels that display information about the application's current state. 
The first panel in Figure \ref{fig:grafanadashboard:image} shows the total number of open connections, while Figure \ref{fig:grafanadashboard:logqlquery} displays the LogQL query required to generate this metric.
To construct the query, developers must be familiar with \texttt{label\_names} and \texttt{label\_values}, as well as the exact log line syntax, including elements such as ``\texttt{sshd[}'' and ``\texttt{: session opened for}'', with precise attention to whitespace formatting. 
These components are human-generated without standardized guidelines, 
which makes it harder for developers to write \logql queries, as the developers do not have complete knowledge of the various log formats, keywords, labels and other components required to compose these queries.

The lack of standardization extends beyond individual log lines to the various log line formats that developers must handle. 
Current log parsers, including LLMParser \cite{llmparser2024ma} and Drain \cite{Drain2017He}, achieve between 50\% to 90\% parsing accuracy across applications, presenting an ongoing challenge in the field.
The absence of a standardized approach for log line composition results in temporal shifts in formats and syntactical elements required for query construction. 
These changes in log line structure create difficulties for developers, even those proficient in query languages, as they attempt to formulate and maintain queries over time.

Since various 
developers face challenges in composing the query, there is a need for a standard query interface that enables developers to write queries more easily.
Given the recent advancements in Large Language Models (LLMs), specifically code generation models for languages like SQL~\cite{texttosql2022pi, text2sqleval2022Rajkumar}, we propose that a Natural Language interface will allow developers easy access to observability data by providing a natural language
interface
over the underlying query language~\cite{sqlauthoring2024Maddila}. 
Generating log queries with an LLM is non-trivial as there are more than 50 different platform %
specific log querying languages, and each model offers varying degree of support for them. 
LLMs require a fundamental understanding of the target query language to generate effective queries. 
Without this knowledge, LLMs may either refuse to generate queries or, in more problematic cases, produce semantically or syntactically incorrect queries~\cite{multilangcode2024Hou}.
The LLMs
often end up generating 
queries that are non-executable and in situations where the LLMs know the query language, efficiently prompting them is often not enough to generate realistic queries (as we demonstrate in~\S\ref{sec:problem_statement:llmassisted}) due to insufficient knowledge of log lines which are too big to be fit into context window of the application.
On the other hand, finetuning the LLMs for various languages requires rich set of natural language questions, corresponding queries and the output from the queries to serve as ground truth to test the efficiency of the model.

The challenge of querying logs with natural language exhibits significant similarities to problems encountered in data analysis, such as text-to-SQL conversion. 
Composing queries for log searches frequently involves the utilization of both structured elements, such as labels~\cite{splunkindexer} and tags~\cite{datadogindexer}, as well as unstructured components from log entries that need to be included or excluded. 
In both scenarios, there is often a disconnect between the individuals generating the data (\eg tables, rows, logs) and those querying it for information extraction~\cite{he2022empirical}. However, these problems diverge in terms of data characteristics and scale.
While SQL tables may have varying schemas, the diversity in log line formats is considerably more extensive. 
For example, common text-to-SQL benchmarks like SPIDER~\cite{spider12018tao} contain 5 tables per database, whereas each application in LogHub~\cite{Loghub2023Zhu} contains more than 25 log formats per application.

In this work, we take the first steps towards building realistic NL interfaces for generating log queries. 
To achieve this, we first create a dataset of natural language to \logql queries and use it to fine-tune a suite of popular LLMs creating \sysName, a system designed to convert natural language questions into \logql queries.
LogQL is a specialized query language for searching and analyzing log data within Grafana's open-source log aggregation system, Loki~\cite{grafana_loki}.
The selection of LogQL was motivated by its open-source nature and the extensive availability of Grafana dashboards (Fig \ref{fig:grafanadashboard:image}), which enable the formulation of realistic natural language queries, and its support in various open source LLMs. 

Through our exhaustive evaluation, we demonstrate that fine-tuning on our dataset significantly enhances the performance of popular LLMs, including GPT-4o, Llama 3.1, and Gemini, in generating accurate LogQL queries. Our experiments reveal that GPT-4o achieves up to 75\% and 80\% improvements in accuracy and F1 respectively, with fine-tuning enhancing query outputs,
reducing syntax errors, and improving label matching and temporal aggregation. We perform further ablation studies to evaluate the effects of the number of training examples and the potential transferability of these models across applications. 

Specifically, this paper makes the following contributions: \vspace{-0.2cm}
\begin{itemize}
\item First, we present and release a dataset—\textbf{NL2LogQL}—designed to facilitate the development and benchmarking of natural language to LogQL systems, with a particular focus on fine-tuning Large Language Models (LLMs) to generate syntactically and semantically correct \logql queries. NL2LogQL consists of 424 manually curated natural language to LogQL pairs. Each pair is derived from a panel in a Grafana Community Dashboard, covering three distinct applications. The dataset was constructed by manually describing the purpose of each panel in natural language and crafting the corresponding \logql query. This resource represents the first dataset specifically designed to enable an NL-to-\logql interface.

\item Second, we present a web-based interface that enables developers to generate LogQL queries for the aforementioned applications. This interface serves a dual purpose: it provides a practical tool for query generation and acts as a platform for collecting additional natural language questions, thereby facilitating the continuous expansion of the dataset.

\item Third, utilizing this novel dataset, we fine-tune three off-the-shelf LLMs for the task of natural language to \logql query generation. We release these fine-tuned NL-to-\logql models, along with the prompts used for their evaluation and a fine-tuned CodeBERTScore model for assessing the results.

\item We establish a set of metrics to quantitatively assess both the syntactic and semantic correctness of \logql queries generated by LLMs, and further conduct a comprehensive study on the efficacy of fine-tuning models, analyzing the optimal number of samples, the impact of post-fine-tuning prompting, and the transferability of models across various applications.

\end{itemize}

\section{NL Interface for Log Search}
\label{sec:problem_statement}

In this section, we present the challenges associated with querying logs, highlight specifics of an open-source log query language called \logql, and present an initial sketch of how we can use LLMs to translate natural language queries to \logql.

\vspace{-0.3cm}
\subsection{Challenge: Querying Logs is Difficult}

Existing tools for storing and querying log data to obtain insights present a significant usability challenges for many reasons. We highlight these challenges below. 

\textbf{Steep Learning Curve}. These tools require the developers to learn and use esoteric tool-specific query languages, that have a steep learning curve~\cite{cncfdsls}. Due to the unintuitive nature of these languages, users often struggle with constructing effective queries to find specific log entries~\cite{honeycomb2}. Consequently, only a small percentage of power users within an organization can leverage the full capabilities of log analysis tools~\cite{honeycomb1}, hindering the democratization of log analysis.
Recent studies~\cite{seeit2022davidson} have also highlighted the challenges faced by new team members in using existing tools, requiring significant time to gain proficiency. 

\textbf{Insufficient Context}. In the context of DevOps \cite{devops2019Wiedemann, mismatches2022seshagiri}, where the developer and the operator are usually different individuals thus the person writing queries often lacks sufficient context. 
To construct effective queries, the operators often require detailed knowledge of log lines. Operators often lack appropriate context as they're dealing with unfamiliar logs, or context switching between multiple tools and dashboards~\cite{he2022empirical,qualistudy2023davidson}.

\textbf{Large and unstructured logs}. The complexity of writing these queries increases due to the high volume of application logs and their varying formats.
For example, to write the LogQL query in Fig.~\ref{fig:grafanadashboard:logqlquery}, the developer must know syntax of the log querying language (\logql) and also the semantics the log file -- such as ``\textcolor{darkgreen}{\emph{sshd[}}'', ``\textcolor{darkgreen}{\emph{session opened for}}'' -- that are required to construct this query. 

Due to these inherent complexities, log data analysis remains predominantly within the domain of software developers who possess intricate knowledge of the logging systems. 
The utilization of log data presents significant untapped potential for informing strategic business decisions~\cite{bizintel2024logzio}.
For example, insights from HTTP header parsing can optimize marketing campaigns by understanding regional traffic patterns, and analysis of distributed traces can reveal the most commonly used platform features.
Questions like ``\emph{what is the average response time for my website?}'' or ``\emph{what are the most common errors being reported?}'' can provide a more detailed picture of system performance and identify areas for improvement.
However, similar to data analysis domain~\cite{nl4dv2020narenchania}, non-engineers who want to extract such information must either rely on engineers to write queries or learn the querying language themselves, creating a barrier to data-driven decision-making across the organization.

\cemph{To write effective log search queries, developers often need to have complete syntactical and semantic knowledge of the query language and log lines. These challenges collectively underscore the need for more intuitive and user-friendly interfaces to log analysis that can improve productivity, accessibility, and cross-functional utility of log data.}

\vspace{-0.2cm}
\subsection{Background: LogQL}
\label{sec:problem_statement:logql}
LogQL is a query language designed for searching and analyzing log data in Grafana's log aggregation system, Loki~\cite{grafana_loki}. Loki focuses on indexing metadata rather than full log text. 
LogQL provides tools for filtering, aggregating, and extracting insights from log streams, supporting both log and metric queries, with a structure that includes label selectors, line filters, and time range specifications.

LogQL provides users with tools to filter, aggregate, and extract insights from log streams, making it valuable for monitoring, troubleshooting, and maintaining complex distributed systems.
Operators typically write a LogQL query per panel that are then arranged together to form a dashboard (for example, the dashboard in Figure \ref{fig:grafanadashboard:image}). 
The language supports two primary query types: log queries for retrieving and filtering log content, and metric queries for applying aggregation functions to transform log data into numerical time series. 

As Loki's indexing strategy focuses on the metadata associated with log lines, rather than the full log text, this indexing approach has implications for LogQL queries, as they must include the relevant tags to search the indexed metadata effectively. 
For example, given the log line [\textsc{2019-12-11T10:01:02.123456789Z \{app="nginx",cluster="us-west1"\} GET /about}], Loki will index the timestamp and the labels attached to the log line, such as ``app'' and ``cluster'', but not the actual log text starting from ``GET''.
Since the indexing of the logs is based on timestamp, the queries are relative to the current system time. 
Thus, LogQL provides labels to allow filtering log lines using metadata. Labels are key-value pairs associated with log streams, providing metadata about the logs' origin and characteristics.

Consider the human written query in Figure \ref{fig:querylang:logql} (green colour) to quantify the occurrence of authentication-related service unavailability errors in an OpenStack deployment within the Asia-Pacific region over the past 30 days.
It begins with label selectors enclosed in curly braces: \emph{\{job=``openstack'', region=``asia-pacific''\}}. 
The \emph{job=``openstack''} label identifies logs from OpenStack services, while \emph{region="asia-pacific"} narrows the focus to the Asia-Pacific region. 
Following the label selectors are two line filters: \emph{|= ``503''} and \emph{|= ``token validation''}. These filters use the \emph{|=} operator to perform case-sensitive matches, selecting log lines containing the HTTP status code 503 (indicating a service unavailable error) and mentioning ``token validation''. 
The query concludes with a time range specification ``\emph{[30d]}'', which defines a 30-day analysis window. 
The ``\emph{$count\_over\_time()$}'' function wraps the entire log selection criteria, counting the number of matching log lines within the specified time range.
The complete list of filters and aggregation commands can be found in LogQL's documentation~\cite{logql_documentation}.

\subsection{Our Vision: LLM assisted query generation}
\label{sec:problem_statement:llmassisted}
The heterogeneity of log query languages necessitates enhanced query composition interfaces. 
While existing ``query builder'' interfaces ensure syntactic correctness, they depend on developers' expertise in log line selection. LLMs have shown efficacy in log-related tasks, leveraging their ability to process unstructured text. 
However, direct LLM application to log search presents challenges in handling large-scale data, efficient indexing, and real-time capabilities \cite{honeycomb2}. 

We propose utilizing LLMs for log query generation, balancing accessibility and efficiency. 
This approach, applied successfully in SQL generation \cite{sqlauthoring2024Maddila} and data analysis \cite{autodvllm2024wu}, leverages LLMs' natural language understanding to translate search intents into optimized queries. 
This method bypasses complex indexing requirements while maintaining LLM capabilities, enabling execution through existing log search systems. 
It reduces the query language learning curve, facilitates faster iteration for experienced developers, and aligns with engineers' mental models of their systems.

\lstdefinestyle{style1}{
    language=SQL,
    backgroundcolor=\color{color1!30},
    basicstyle=\ttfamily,
    frame=single
}

\lstdefinestyle{style2}{
    language=SQL,
    backgroundcolor=\color{color2},
    basicstyle=\ttfamily,
    frame=single
}

\lstdefinestyle{style3}{
    language=SQL,
    backgroundcolor=\color{color3},
    basicstyle=\ttfamily,
    frame=single
}

\lstdefinestyle{style4}{
    language=SQL,
    backgroundcolor=\color{color4},
    basicstyle=\ttfamily,
    frame=single
}

\begin{figure}[ht]
\begin{subfigure}{\linewidth}
\begin{mdframed}[   
linecolor=blue!70!black,
backgroundcolor=color1!10,
linewidth=0pt,
roundcorner=10pt,
innerleftmargin=10pt,
innerrightmargin=10pt,
innertopmargin=6pt,
innerbottommargin=6pt,]
\begin{lstlisting}[
language=SQL,
basicstyle=\ttfamily\small,
]
How many times did we receive a 503 status 
code while validating tokens in the past 
30 days for openstack-asia-pacific?
\end{lstlisting}
\end{mdframed}
\caption{Summarized NL Query, complete NL query in our code repository}
\label{fig:querylang:nlquery}
\end{subfigure}

\begin{subfigure}{\linewidth}
\begin{mdframed}[   
linecolor=blue!70!black,
backgroundcolor=color3,
linewidth=0pt,
roundcorner=10pt,
innerleftmargin=10pt,
innerrightmargin=10pt,
innertopmargin=6pt,
innerbottommargin=6pt,]

\begin{lstlisting}[
language=SQL,
basicstyle=\ttfamily\small\color{darkgreen},
]
count_over_time({job="openstack", region
="asia-pacific"} |= "503" |= "token 
validation" [30d])
\end{lstlisting}

\begin{lstlisting}[
language=SQL,
basicstyle=\ttfamily\small\color{red}
]
calculate_over_time({job="openstack", 
region="asia-pacific"} |= "503" != 
"token" [30d])
\end{lstlisting}
    
\end{mdframed}
\caption{LogQL Query (Green: Correct, Red: LLM Generated)}
\label{fig:querylang:logql}
\end{subfigure}

\begin{subfigure}{\linewidth}
\begin{mdframed}[   
linecolor=blue!70!black,
backgroundcolor=color2,
linewidth=0pt,
roundcorner=10pt,
innerleftmargin=10pt,
innerrightmargin=10pt,
innertopmargin=6pt,
innerbottommargin=6pt,]
\begin{lstlisting}[
language=SQL,
basicstyle=\ttfamily\small\color{darkgreen}
]
@http.status_code:503 service:openstack
-asia-pacific "validate" @timestamp:>now-30d
\end{lstlisting}

\begin{lstlisting}[
language=SQL,
basicstyle=\ttfamily\small\color{red}
]
status.code:503 service_name:openstack-asia-pacific "validate" @timestamp:>now-30d
\end{lstlisting}
    
\end{mdframed}
\caption{Datadog Query (Green: Correct, Red: LLM Generated)}
\label{fig:querylang:dd}
\end{subfigure}

\begin{subfigure}{\linewidth}
\begin{mdframed}[   
linecolor=blue!70!black,
backgroundcolor=color4,
linewidth=0pt,
roundcorner=10pt,
innerleftmargin=10pt,
innerrightmargin=10pt,
innertopmargin=6pt,
innerbottommargin=6pt,]

\begin{lstlisting}[
language=SQL,
basicstyle=\ttfamily\small\color{darkgreen},
]
grep -E "status: 503.*validate_token" OpenStack_2k.log | grep "openstack-asia-pacific" | wc -l
\end{lstlisting}

\begin{lstlisting}[
language=SQL, 
basicstyle=\ttfamily\small\color{red}
]
grep "ERROR 503" OpenStack.log | awk '{print $4}' | sort | uniq -c
\end{lstlisting}

\end{mdframed}
\caption{\texttt{grep}-Based Query (Green: Correct, Red: LLM Generated)}
\label{fig:querylang:grep}
\end{subfigure}

\caption{Queries for analyzing 503 status codes in OpenStack Asia-Pacific %
across different query languages}
\label{fig:querylang}
\end{figure}

To adapt LLMs for specific tasks, two lines of approaches have been  employed in the past: (i) In-Context Learning (ICL)~\cite{brown2020languagemodelsfewshotlearners}; and (ii) fine-tuning pre-trained models with task-specific examples. We discuss the potential and limitations of both in generating LogQL queries below.

\textbf{ICL} incorporates task-specific demonstrations into the input during inference, guiding the model without parameter retraining. However, incorporating large log files is impractical for smaller models due to limited context windows~\cite{honeycomb3}, and models with larger context windows often exhibit instability and reduced robustness~\cite{lostinmiddle2024Liu}.
Similar to SQL generation issues \cite{Nextgendbinterface2024Hong}, LLMs often generate non-existent log lines. 
To demonstrate this, we devised a prompt 
with documentation, examples, and instructions for generating queries in Datadog Query Language (DQL), LogQL, and \texttt{grep}.

Figure \ref{fig:querylang} presents an example query for searching service not found errors during token validation in the past 30 days for a specific OpenStack node. 
Comparison of LLM-generated queries (red color) with human-crafted (green color) ones reveals semantic inconsistencies across DQL, LogQL, and \texttt{grep} methods. 
In DQL, the LLM-generated query uses non-standard attribute names (e.g., ``status.code:503'' instead of ``@http.status\_code:503''), indicating gaps in platform-specific convention comprehension. 
The LogQL LLM query erroneously employs a non-existent function (e.g., ``calculate\_over\_time'' instead of ``count\_over\_time'') and misapplies operators (using ``!='' instead of ``|='' for inclusion). 
The \texttt{grep} LLM query presents a generic structure, lacking context-specific knowledge of log formats (using generic error statements like ``error'' or ``failed'' instead of specific log patterns) and misapplying common log analysis patterns (e.g., searching for ``token validation'' instead of specific token-related log entries).
These errors, combined with log lines not fitting into the context window, demonstrate that using ICL based approaches fails to generate realistic log query language queries.

\textbf{Fine-tuning}, particularly few-shot tuning, offers significant advantages for adapting pre-trained LLMs to specific tasks such as LogQL generation. 
This approach involves re-training the LLM on a tailored dataset, allowing the model to adjust its internal parameters and better align its outputs with desired outcomes. 
Few-shot tuning enables LLMs to generalize from limited examples, facilitating the extraction of relevant information across diverse log formats and applications. 
This is particularly crucial given the often \emph{ad hoc} nature of log files, which lack standardized logging procedures. 
By providing more diverse log examples during the training phase, fine-tuning enhances the model's ability to handle varied log structures. 

Prior studies \cite{llmfinetuning2024sarker, sense2024yang} have demonstrated that few-shot tuning offers superior accuracy at lower computational costs for related tasks like text-to-SQL. 
Moreover, the efficiency of few-shot tuning, requiring only a small number of data samples, results in a rapid fine-tuning process without significant time overhead. 

Importantly, few-shot tuning eliminates the need for continuous in-context demonstrations during inference, potentially reducing overall query latency—a critical bottleneck for log queries \cite{odms2021Karumuri}. 
This reduction in query latency is especially vital in log search systems, where rapid data retrieval and analysis are essential for real-time monitoring and troubleshooting of complex distributed systems. 
Faster query times enable IT teams to detect and respond to issues more quickly, minimizing downtime and improving overall system performance.

\cemph{Few-shot tuning offers a promising approach for adapting LLMs to log query generation tasks. 
This method enables efficient generalization from limited examples, reduces inference time, and allows for diverse log example training. 
Given these advantages, we employ few-shot tuning to fine-tune LLMs for LogQL query generation in this study.}

\section{\sysName{}}
\label{sec:system_new}

We first define the problem of translating a natural language log-query to \logql as follows.

\begin{definition}[NL2\logql]
\label{def:nl2logql}
Let $DB$ be an unstructured logfile, let $q^{NL}$ be a query specified in natural language, and let $q^{LOG}$ be the corresponding query expressed in LogQL, that when executed against $DB$ can provide (semantically correct) intended query answer $a$.

We define the $NL2\logql$ problem as finding a mapping of any natural language query $q$ to a syntactically and semantically correct \logql query:

$$
NL2\logql(q^{NL}) \mapsto q^{LOG}
$$
Such a query can then be executed against $DB$ directly to produce the intended answer.
$$
q^{LOG}(DB) \rightarrow a
$$
\end{definition}

To train such a mapping $NL2\logql$, and evaluate its logging efficacy, we first require a dataset that provides us with example tuples $(DB,q^{\text{NL}},q^{LOG}, a)$ each having a logfile $DB$, a natural language query $q^{\text{NL}}$, a ground truth correct LogQL query $q^{\text{LOG}}$, and the output of executing the query, $a$. 
We manually create such a dataset having 424 example tuples as described in Section~\ref{sec:system_new:dataset}. Using this dataset, we describe our approach to fine-tuning existing large language models to automatically map any new natural language query for any new logfile to the intended \logql query.

\subsection{Dataset}
\label{sec:system_new:dataset}
As defined earlier, to effectively fine-tune a model to tranform natural language queries to their LogQL counterparts, we require a comprehensive dataset consisting of $(DB, q^{\text{NL}}, q^{\text{LOG}}, a)$ tuples. Table~\ref{tab:example_dataset} shows examples of such tuples for two metric queries and two log queries.

\begin{table*}[t]
\begin{tabular}{|l|l|l|}
\hline
\textbf{NL Query} & \textbf{LogQL} & \textbf{Query Output} \\ \hline
\makecell[l]{How many times did the NameSystem \\allocate new blocks in the past minute for \\hdfs-south-america?} & \makecell[l]{sum(count\_over\_time(\{application=\\``hdfs-south-america''\}|$\sim$``BLOCK\textbackslash{}\textbackslash{}* \\NameSystem\textbackslash{}\textbackslash{}.allocateBlock:'' {[}1m{]}))} & 1880\\ \hline
\makecell[l]{How many times did PAM ignore max \\retries in the last 24 hours \\for openssh-us-east?} & \makecell[l]{sum(count\_over\_time(\{application="openssh", \\hostname="us-east"\} |= "PAM service(sshd) \\ignoring max retries" {[}24h{]}))} & 39700 \\ \hline
\makecell[l]{Show me the most recent successful login \\for user 'fztu' in openssh-asia-pacific,\\ including timestamp and source IP?} & \makecell[l]{\{application="openssh-asia-pacific"\} \\|= "Accepted password for fztu" | regexp \\"(?P\textless{}source\_ip\textgreater{}\textbackslash{}\textbackslash{}d+\textbackslash{}\textbackslash{}.\textbackslash{}\textbackslash{}d+\textbackslash{}\textbackslash{}.\textbackslash{}\textbackslash{}d+\textbackslash{}\textbackslash{}.\textbackslash{}\textbackslash{}d+)"
} & \makecell[l]{120 Log lines with \\ Accepted password \\ for fztu} \\ \hline
\makecell[l]{What are the top 3 most frequent \\exceptions encountered during writeBlock \\ operations in the past 24 hours for \\ hdfs-asia-pacific? }& \makecell[l]{topk(3, sum by (exception\_type) ( count\_over\\\_time(\{component=$\sim$"dfs.DataNode.*", application=\\"hdfs-asia-pacific"\} |$\sim$"writeBlock .* received exception" \\| regexp "writeBlock .* received exception \\(?P\textless{}exception\_type\textgreater{}{[}\textasciicircum{}:{]}+)" {[}24h{]})))}& \makecell[l]{\{exception\_type="java.io.\\EOFException"\} \\
\{exception\_type="java.io.\\IOException"\} \\
\{exception\_type="java.io.\\InterruptedIOException"\}}\\ \hline
\end{tabular}\vspace{0.2cm}
\caption{Example tuple from our dataset showing the NL query
\logql query and the corresponding output. The first 2 rows represent metric queries and the next 2 represent log queries}
\label{tab:example_dataset}
\end{table*}

\textbf{Data Sources}. Constructing a dataset of such records  necessitates both realistic logs and natural language questions to ensure the model's applicability for operators writing queries on their applications. 
We source logs from the LogHub 2.0 dataset~\cite{Loghub2023Zhu}, which encompasses logs from diverse applications and includes various event extraction templates for log parsing tasks. 
To obtain realistic NL queries, we analyze the Grafana Community Dashboards \cite{grafana_community_dashboards}, as these dashboards are open source and publicly available. We extract NL questions based on panel titles and displayed information.
Our analysis of these dashboards informs the dataset construction, involving the creation of \logql queries corresponding to the presented panels. 
For panels requiring multi-step queries, such as pie charts displaying failed login attempts by users, we decompose them into separate NL to \logql entries in our dataset.
For example, \autoref{fig:grafanadashboard:image} illustrates an example dashboard for OpenSSH, publicly available and comprising 7 metrics panels and 2 log panels, which provides valuable information for developers to establish alerting systems or query data. (e.g., ``show the total users with failed attempts'' and ``how many failed login attempts for $\$username$'').
While we utilize Grafana and \logql for our dataset construction, our approach is extensible to other dashboards and log query languages.

Drawing insights from the construction of analogous datasets for text-to-SQL or data analysis tasks \cite{sense2024yang, SyntheticSQL2024Meyer, Sofset2024Barke}, we recognize that dataset diversity is crucial for producing fine-tuned models capable of addressing a wide range of queries. 
To develop models that are valuable for operators in querying their logs, we have identified three key domains across which we ensure dataset diversity: application type, use case, and \logql operation complexity. 
We create a total of 424 individual entries in our dataset, across 3 applications encompassing a wide range of use cases and \logql operations. 
We spent 500 hours of human labour between the authors to create and validate the dataset. 

\textbf{Application Diversity} Our dataset mirrors the database diversity observed in text-to-SQL tasks, where the models are finetuned to generate queries across different databases. 
This diversity is essential for log analysis, as each application generates unique log formats, lines, and structures, which are crucial for constructing accurate \logql queries.
To ensure a representative range, we have built our dataset using logs and dashboards from three distinct applications: OpenSSH, OpenStack, and HDFS. 
These applications span diverse domains of system operations: OpenSSH facilitates secure network communications, OpenStack manages cloud computing resources, and HDFS enables distributed storage of large data volumes across commodity hardware clusters.
For the dataset, we wanted to have at least 50 samples per application, to be consistent with common text-to-SQL benchmarks such as Spider\cite{spider12018tao} and BIRD\cite{bird2024li}.
Our dataset comprises 155, 154, and 115 samples for OpenSSH, OpenStack, and HDFS, respectively.

\textbf{Use case Diversity} 
For OpenSSH, we identified 7 distinct use cases: Suspicious Activities, Brute Force Attempts, Connection Analysis, Invalid User Attempts, System Health and Performance, User Session Analysis, and Authentication Failures. 
OpenStack presented 11 use cases, including Instance Lifecycle, Audit and Synchronization, Resource Usage, System Health and Maintenance, API Performance, Instance Lifecycle Management, Image and File Management, Network Operations, Security and Authentication, Error Analysis, and API Performance and Requests. 
HDFS contributed 7 use cases: Replication and Data Transfer, Error Analysis, Performance Issues, Data Transfer and Replication, Performance Monitoring, Block Management, and NameNode Operations. 
This variety of use cases across applications ensures that our fine-tuned models can address a wide spectrum of log analysis scenarios, enhancing their practical utility for operators.
{\large
\begin{table}[t]
\begin{tabular}{@{}|l|l|l|@{}}
\hline
Type of query           & Filter(s)      & Percentage \\ \hline
\multirow{4}{*}{Log}    & Single Line    & 65.8 \\ \cline{2-3} 
                        & Multiple Line  & 34.2 \\ \cline{2-3} 
                        & Single Label   & 36.5 \\ \cline{2-3} 
                        & Multiple Label & 63.5 \\ \hline
\multirow{3}{*}{Metric} & Log RA         & 40.1 \\ \cline{2-3} 
                        & Unwrapped RA   & 7.8  \\ \cline{2-3} 
                        & Built-in RA    & 40.1 \\ \hline
\end{tabular}
\caption{LogQL Query Types and Filters with corresponding values in our dataset\vspace{-1cm}}
\label{tab:datasetcomposition}
\end{table}
}

\textbf{\logql Operation Diversity} The composition of our dataset reflects this diversity in \logql operations, as illustrated in Table \ref{tab:datasetcomposition}. 
For log queries, 65.8\% use single line filters, while 34.2\% employ multiple line filters. In terms of label filters, 36.5\% of queries use single label filters, and 63.5\% use multiple label filters. 
It is important to note that these percentages are independent; a query with a single label filter can still have multiple line filters, and vice versa. 
This distribution ensures a balanced representation of both simple and complex log query structures. 
For metric queries, we observe an equal distribution between log range aggregation and built-in range aggregation, each accounting for 40.1\% of the metric queries.
Unwrapped range aggregation is less common but still represented, comprising 7.8\% of the metric queries. 
This distribution of query types and filters in our dataset provides a robust foundation for finetuning models capable of handling a wide array of \logql query scenarios.

\subsection{Finetuning LLMs}
\label{sec:system_new:finetuning_llms}
In this section, we present the models used for fine-tuning and prompting for the task of~\textit{NL2\logql}. 

For finetuning, we require models that can excel at various coding and reasoning tasks and can learn to specifically generate syntactically and semantically correct \logql queries.

To ensure a systematic empirical evaluation across a diverse set of models, we selected three widely recognized models that have achieved state-of-the-art performance in various coding and natural language tasks. Specifically, we employed LLama-3.1~\cite{llama2023touvron2023}, Gemma-2~\cite{team2024gemma}, and GPT4o~\cite{openaigpt4}, for subsequent fine-tuning of~\textit{NL2QL}. Each model has unique strengths, especially in reasoning-heavy tasks like Measuring Massive Multitask Language Understanding (MMLU)
and coding benchmarks~\cite{hendrycksmeasuring}.
\begin{itemize}
    \item~\texttt{GPT4o} is robust in both reasoning and coding, especially for complex queries, but it's closed-source and proprietary.
    \item~\texttt{Gemma-2-9B} is the smallest LLM of the Gemma-2 series, released by Google, which, despite its size, offers competitive performance due to improved parallelized training. This model balances efficiency and performance, making it ideal for scalable applications like~\textit{NL2\logql}.
    \item~\texttt{LLama-3.1-8B} is another notable LLM, leveraging Grouped-Query Attention (GQA) for improved inference scalability. GQA enables more efficient handling of large input sequences, which is critical for inference.
\end{itemize}
We fine-tune these models using LoRA~\cite{hulora}, which introduces a small set of additional trainable parameters while freezing the original model. By using low-rank parameterization, LoRA reduces computational and memory costs, enabling faster convergence with minimal performance impact. 
\textbf{Training Hyperparameters}: We used the AdamW optimizer~\cite{loshchilov2018decoupled} with 8-bit quantization, implemented via the bitsandbytes library. The learning rate was set to 1e-4, following a cosine schedule with a warm-up of 10 steps. The training was run for 4 epochs with a micro-batch size of 4 and gradient accumulation steps of 1. To optimize memory usage, we applied gradient checkpointing, 8-bit quantization for the base model, and flash attention. LoRA parameters include a rank of 16, a scaling factor of 32, and a dropout rate of 0.05. All our models were trained on 1 Nvidia A100 GPU available via Modal Labs~\cite{modallabs}

\subsection{Metrics}
\label{sec:system_new:metrics}
Similar to prior work in text-to-SQL~\cite{picard2021scholak}, we evaluate the performance of the our finetuned models using Exact Match and Execution Accuracy. 
To assess Execution Accuracy, we compared the results returned by the \logql queries generated by our models to the results from manually written reference \logql queries. 
Since there are two main types of \logql queries, Metric and Log, we used different metrics tailored to each. 
\begin{itemize}
    \item For \textbf{Metric queries}, which return a numerical value, we compared the model's output and expected output. 
    We find the output accurate if it is exactly the same as the expected output, and any deviation is considered a wrong output, for floating point outputs we compare up to two decimal places rounded up. 
    \item For \textbf{Log queries}, which return a list of relevant log lines, we computed the precision and recall of the model's output log lines compared to the reference log lines. 
    We calculate the F1 score, which is the harmonic mean of precision and recall, as a summary metric. 
    These metrics evaluate how well the model's queries are filtering the logs to surface the most pertinent information.
    Since \logql results are always sorted by timestamp, we did not need to compare the relative ranking of the model and reference outputs - the ordering is guaranteed to be consistent as long as the same logs are returned.
\end{itemize}
Along with these metrics, we also make use of Perplexity Score~\cite{Perplexity2024Xu}, to assess the language model's ability to predict the next token in a sequence, providing a measure of how well the generated code aligns with the model's learned probability distribution. 
A lower perplexity score indicates that the model is more confident and accurate in its code predictions. 

While comparing the output of the model is meaningful for checking the end result, it doesn't account for the syntax lapses causing the outputs to be dramatically different. 
Relying solely on output-based evaluation methods can be misleading, as they may fail to capture the underlying issues in the generated \logql queries. 
To address these limitations, we assess the exact match between the generated query to ground truth query. 

While Perplexity scores offers an intrinsic metric for the confidence of a model in generating accurate \logql queries, it doesn't provide an extrinsic metric. 
To mitigate this challenge, we make use of CodeBERTScore(CBS)~\cite{codebertscore2023Zhou}.
CBS is a pre-trained language model specifically trained
for evaluating code outputs in various programming languages. 
By finetuning CBS on a subset of \logql queries from our dataset, we evaluate the generated \logql queries based on their semantic similarity to reference queries and their adherence to the syntactic rules of the \logql language.
The CBS evaluates the similarity between generated and reference \logql queries using cosine similarity, producing values between 0 and 1. 
A score of 1 indicates perfect semantic and functional equivalence, while 0 represents complete dissimilarity. 
Scores above 0.7 strongly correlate with high query quality and correctness as judged by human evaluators, whereas scores below 0.4 typically indicate significant functional deficiencies. 
The metric's relative nature means it is most meaningful when comparing queries within the same evaluation context.

By combining Perplexity Score and CodeBERTScore, we can obtain a more comprehensive assessment of the generated \logql queries.
These metrics provide insights into the model's ability to generate syntactically correct and semantically meaningful code, which is essential for understanding the model's performance and limitations. 

\subsection{Demonstration}
\begin{figure}[t]
\centering
\includegraphics[width=80mm]{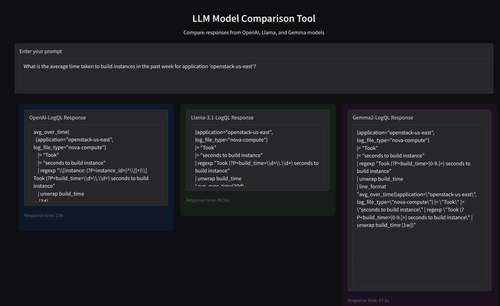}
     \vspace{-0.5cm}
\caption{Demonstration of the model.\vspace{-0.6cm}} 
\label{fig:demonstraion}
\end{figure}

To facilitate comprehensive model comparison and enhance user interaction, we developed a web-based interface that enables simultaneous evaluation of multiple LLM responses. 
As illustrated in Figure \ref{fig:demonstraion}, the interface presents a streamlined design with a prominent query input field at the top, where users can formulate natural language questions about log analysis.
Upon submission, the system concurrently processes the query through three distinct fine-tuned models: GPT-4o, Llama-3.1, and Gemma2. 
The responses are displayed in parallel panels, each showcasing the generated LogQL query along with its response time. 
For example, when querying about instance build times in OpenStack deployments, each model generates specialized LogQL syntax incorporating regex patterns and temporal aggregations. 
The interface displays response times (2.9s, 49.56s, and 67.8s respectively), enabling quantitative performance comparison. 
This parallel visualization approach not only facilitates direct comparison of query formulation strategies but also provides valuable feedback mechanisms for continuous model improvement. 
The comparative layout effectively highlights the nuanced differences in how each model interprets and translates natural language queries into LogQL syntax, contributing to our understanding of model behavior and performance characteristics.
The demo is hosted on \textcolor{violet}{\url{https://llm-response-simulator-alt-glitch.replit.app}}

\section{Evaluation}
\label{sec:evaluation}

To validate our natural language interface for \logql query generation, we established a comprehensive evaluation framework focusing on the models' capability to generate executable queries that yield accurate results.
The evaluation framework covered several key areas: first, a comparison between fine-tuned and baseline models; second, an analysis of how the size of the fine-tuning dataset affects model performance; third, an exploration of how well the model transfers across different application domains; and finally, a qualitative review of the generated \logql queries, using CodeBERTScore as an objective measure. 
This comprehensive approach allowed for a detailed assessment of \sysName's robustness and its ability to generalize to different use cases. 

To ensure compatibility with Loki's indexing system (\S\ref{sec:problem_statement:logql}), we preprocessed the logs by converting them to Loki format using the log parsing templates provided in the LogHub dataset. 
Key-value pairs for the various labels were derived from the keys in the log templates, with corresponding values parsed from each log line using existing parsers. 
Since Loki performs relative-time querying and indexes data based on timestamp, we converted timestamps in logs to to reflect more recent dates, while maintaining relative ordering of the log lines. 
Since most log queries are for searching through the logs in the last 7 days~\cite{odms2021Karumuri}, this pre-processing step proved crucial for facilitating efficient querying and analysis of the log data. 

Natural language questions from our test set are processed through different infrastructures: vLLM for LLama and Gemma models, and OpenAI API for GPT-4o. 
The generated LogQL queries are then executed on a locally deployed Loki instance, ensuring consistent execution conditions and eliminating network-related variabilities.
The responses undergo comparative analysis, measuring various performance metrics as detailed in Section \ref{sec:system_new:metrics}.

\subsection{Performance of finetuned models}
\label{sec:evaluation:baseft}

\begin{table}[ht]
\centering
\resizebox{\columnwidth}{!}{%
\begin{tabular}{|c|@{}|c|l|c|l|c|c|c|@{}|}
\hline
Model                 & App  & MQ (B) & MQ (A) & LQ (B) & LQ (A) & Pplx ($\downarrow$)\\ \hline
\multirow{3}{*}{\raisebox{0.5\height}{\includegraphics[width=0.9cm]{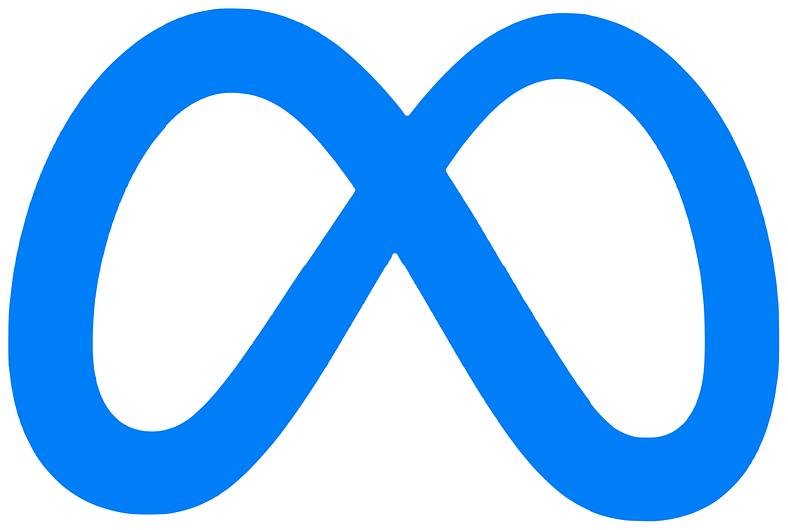}}} \newline Llama & OSSH & 0.03 & 0.50  & 0.05 & 0.42 & 15.2 \\
                           & OSTK & 0.05 & 0.45 & 0.06 & 0.42 &  19.8 \\
                           & HDFS & 0.02 & 0.48 & 0.1 & 0.59  & 22.5 \\ \hline
\multirow{3}{*}{\raisebox{0.5\height}{\includegraphics[width=0.9cm]{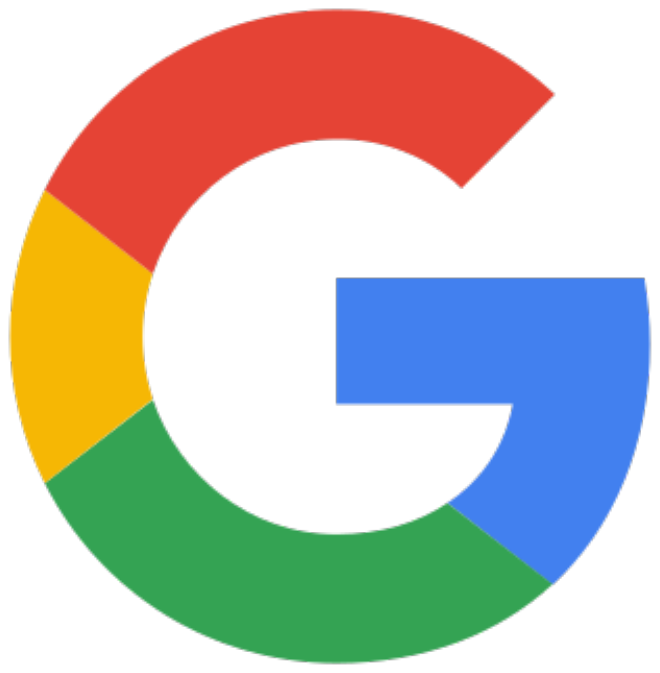}}} \newline Gemma  & OSSH & 0.06 & 0.25 & 0.11 & 0.31 & 38.0 \\ 
                           & OSTK & 0.02 & 0.27 & 0.12 & 0.47 & 36.7 \\ 
                           & HDFS & 0.07 & 0.35 & 0.14 & 0.39 & 27.7 \\ \hline
\multirow{3}{*}{\raisebox{0.5\height}{\includegraphics[width=0.8cm]{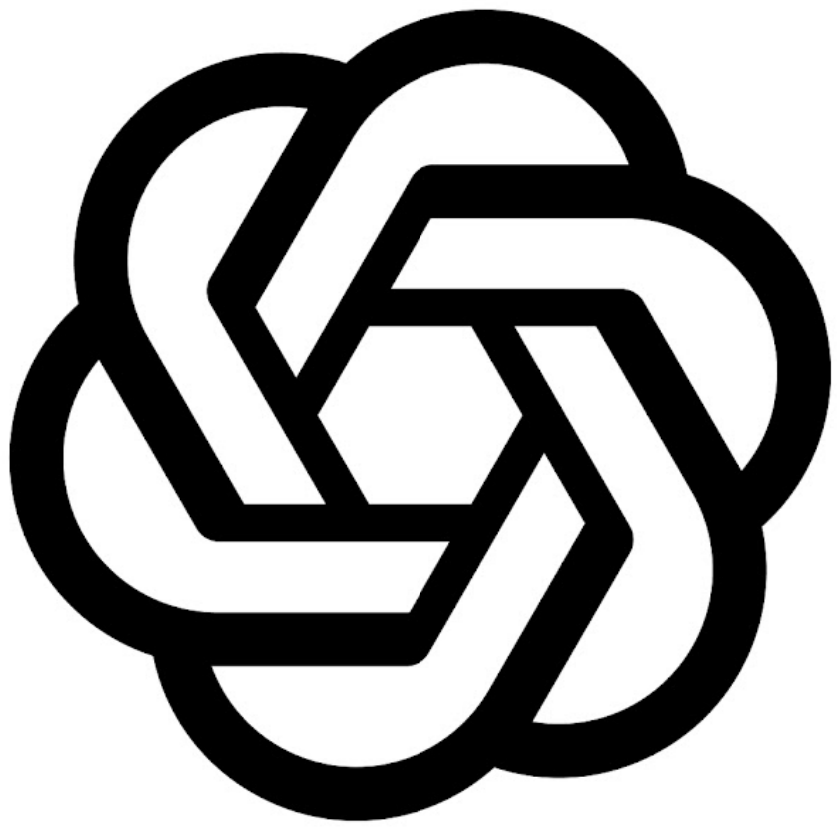}}}  \newline GPT4-o  & OSSH & 0.21 & 0.74 & 0.16 & 0.62 & 9.8  \\
                           & OSTK & 0.28 & 0.82 & 0.18 & 0.68 & 10.2 \\
                           & HDFS & 0.23 & 0.79 & 0.19 & 0.74 & 10.7 \\ \hline
\end{tabular}}
\caption{Results for models (B)efore and (A)fter finetuning. MQ = Metric Queries measured by Accuracy; LQ = Log Queries measured by F-1 Score; Pplx = Perplexity 
}
\label{tab:finetune_model_perf}
\end{table}

In this experiment, we wanted to test the effeciency of the finetuned models in generating \logql queries compared to the base model. 
We used 50\% of the samples from each application for finetuning the models using the method described in Section \ref{sec:system_new:finetuning_llms}.
Our experimental evaluation of the NL2LogQL translation models reveals significant 
improvements in accuracy and F-1
score through finetuning, as illustrated in Table \ref{tab:finetune_model_perf}. 
Most of the queries generated were executable, except for 10\% of the queries that had wrong syntax such as no log lines after filters or had ill-formed regular expressiones. 
Among the finetuned models, GPT-4o exhibited the strongest performance, achieving remarkable post-finetuning accuracy scores ranging from 0.74 to 0.82 and F1 scores between 0.62 and 0.74, up from pre-finetuning metrics of 0.21-0.28 for accuracy and 0.16-0.19 for F1 scores. 
Llama-3.1 showed significant enhancement, with accuracy improving from below 0.05 to approximately 0.50 across applications, alongside F1 scores rising from around 0.05 to the 0.42-0.59 range. 
While Gemma-2 demonstrated more modest gains, it still showed meaningful improvements, with accuracy increasing from below 0.07 to 0.25-0.35 and F1 scores improving from 0.11-0.14 to 0.31-0.47. The perplexity scores further support these findings, with GPT-4o achieving the lowest perplexity (9.8-10.7), followed by Llama-3.1 (15.2-22.5), and Gemma-2 (27.7-38.0), indicating superior model coherence and predictive capability.
Most of the correct responses from the models before finetuning came from providing a lot of context to the model for generating the LogQL query.

\lstdefinestyle{style1}{
    language=SQL,
    backgroundcolor=\color{color1!30},
    basicstyle=\ttfamily,
    frame=single
}

\lstdefinestyle{style2}{
    language=SQL,
    backgroundcolor=\color{color2},
    basicstyle=\ttfamily,
    frame=single
}

\lstdefinestyle{style3}{
    language=SQL,
    backgroundcolor=\color{color3},
    basicstyle=\ttfamily,
    frame=single
}

\lstdefinestyle{style4}{
    language=SQL,
    backgroundcolor=\color{color4},
    basicstyle=\ttfamily,
    frame=single
}

\begin{figure}[h]

\begin{subfigure}{\linewidth}
\begin{mdframed}[   
linecolor=blue!70!black,
backgroundcolor=color3,
linewidth=0pt,
roundcorner=10pt,
innerleftmargin=10pt,
innerrightmargin=10pt,
innertopmargin=6pt,
innerbottommargin=6pt,]

\begin{lstlisting}[
language=SQL,
basicstyle=\ttfamily\small\color{darkgreen}
]
topk(1, sum by (source_ip)(count_over_time( 
{application="hdfs-us-east", component=
"dfs.DataNode$DataTransfer"} \|~ "Transmitted 
block .* to .*" \| regexp "(?P<source_ip>[\\d\\.]+)
:\\d+:Transmitted block .* to .*" [12h])))
\end{lstlisting}

\begin{lstlisting}[
language=SQL,
basicstyle=\ttfamily\small\color{black},
breaklines=false,
commentstyle=\color{red}
]
topk(1, sum (source_ip) ( count_over_time( 
{app="hdfs-us-east", component=
"dfs.DataNode$DataTransfer"} |~ /*[12h]*/
"Transmitted block .* to .*"  | ))
\end{lstlisting}
    
\end{mdframed}
\caption{LogQL query for NL Query to find the node with most number of successfull block transmissions. The wrong query contains time aggregration in the middle of the query. [GPT-4o]}
\label{fig:noftvsftsamples:s1}
\end{subfigure}

\begin{subfigure}{\linewidth}
\begin{mdframed}[   
linecolor=blue!70!black,
backgroundcolor=color3,
linewidth=0pt,
roundcorner=10pt,
innerleftmargin=10pt,
innerrightmargin=10pt,
innertopmargin=6pt,
innerbottommargin=6pt,]

\begin{lstlisting}[
language=SQL,
basicstyle=\ttfamily\small\color{darkgreen}
]
{application="openssh"} |= "Did not receive 
identification string from" | hostname="LabSZ
-tenant-5" | line_format "`{{__timestamp__}}
`- Failed to receive identification string 
from {{.content}}"
\end{lstlisting}

\begin{lstlisting}[
language=SQL,
basicstyle=\ttfamily\small\color{black},
commentstyle=\color{red}
]
{hostname!="LabSZ-tenant-5", /*app ="ssh"*/} 
!= "Did not receive identification string 
from" | line_format "{{/*timestamp*/}} - No 
identification from {{.message}}"/*[1m]*/
\end{lstlisting}

\end{mdframed}
\caption{List of all instances where there was a failure in receiving an identification string from host 'LabSZ-tenant-5'. The wrong query contains wrong label (app), wrong timestamp format and improper time aggregation. [Llama]}
\label{fig:noftvsftsamples:s2}
\end{subfigure}

\begin{subfigure}{\linewidth}
\begin{mdframed}[   
linecolor=blue!70!black,
backgroundcolor=color3,
linewidth=0pt,
roundcorner=10pt,
innerleftmargin=10pt,
innerrightmargin=10pt,
innertopmargin=6pt,
innerbottommargin=6pt,]

\begin{lstlisting}[
language=SQL,
basicstyle=\ttfamily\small\color{darkgreen},
]
sum by (component) ( count_over_time(
{application="openstack-eu-west", component
="nova.virt.libvirt.imagecache"}|~ "Active 
base files: (?P<file_path>/.*)"[1h]))
\end{lstlisting}

\begin{lstlisting}[
language=SQL, 
basicstyle=\ttfamily\small\color{black},
commentstyle=\color{red}
]
sum by (component) ( count_over_time /*by*/ 
(file_path)({application="openstack-eu-west", 
component="nova.virt.libvirt.imagecache"} 
|~ "Active base files: (?P<file_path>/.*)") )
\end{lstlisting}

\end{mdframed}
\caption{LogQL query for retrieving the total size of all base files in openstack-eu-west. This query lacks time aggregation ([1h]) block, wrong syntax of using ``by'' which is present in the correct query. [Gemma]}
\label{fig:noftvsftsamples:s3}
\end{subfigure}

\caption{Examples of logql queries generated by the models before (black color, with errors in red) and after (green color) finetuning.}
\label{fig:noftvsftsamples}
\end{figure}

Figure \ref{fig:noftvsftsamples} contains samples of \logql queries generated by the LLM before and after finetuning the model using our dataset. 
Before finetuning, the LLM generated queries exhibited several common errors: incorrect label usage (e.g., ``app'' instead of ``application''), syntax errors in timestamp placements, wrong filter specifications, invalid grouping syntax, and improper matching operators. 
For instance, in Figure \ref{fig:noftvsftsamples:s1}, the pre-finetuning query for GPT-4 showed incorrect label usage and timestamp placement. 
Similarly, Figure \ref{fig:noftvsftsamples:s2} demonstrates Llama's incorrect use of negative matching operators, while Figure \ref{fig:noftvsftsamples:s3} shows Gemma's invalid grouping syntax in \texttt{count\_over\_time} operations.

The fine-tuned models exhibited substantial enhancement in query generation capabilities, producing syntactically valid and executable \logql queries.
For example, in Figure \ref{fig:noftvsftsamples:s1}, the corrected query properly implements regexp capture groups \newline \verb|((?P<source_ip>[\\d\\.]+))| for IP extraction and uses correct sum aggregation with proper label matchers.
In Figure \ref{fig:noftvsftsamples:s2}, the finetuned model correctly uses positive matching operators (|=) and proper application labeling (application=``openssh''). 
Figure \ref{fig:noftvsftsamples:s3} shows the correct implementation of \texttt{count\_over\_time} with proper time window specification [1h] and appropriate temporal aggregation. 
These improvements resulted in queries that could accurately capture and format the desired log data while maintaining proper syntax and execution capability.

\cemph{The experimental results demonstrate that finetuning significantly enhanced the performance of all three LLM models in generating \logql queries, with GPT-4o showing the most impressive improvements in Accuracy and F1 scores by up to 75\% and 80\% respectively. 
Post-finetuning, the models produced 20\% more
executable queries with fewer syntax errors, improved label matching, and better temporal aggregation, highlighting the effectiveness of the finetuning process in enhancing the models' ability to generate accurate and functional \logql queries.}

\subsection{Effect of number of finetuning samples}
\label{sec:evaluation:ftsamples}

\begin{figure*}
\begin{subfigure}{\linewidth}
\centering
\includegraphics[width=\textwidth]{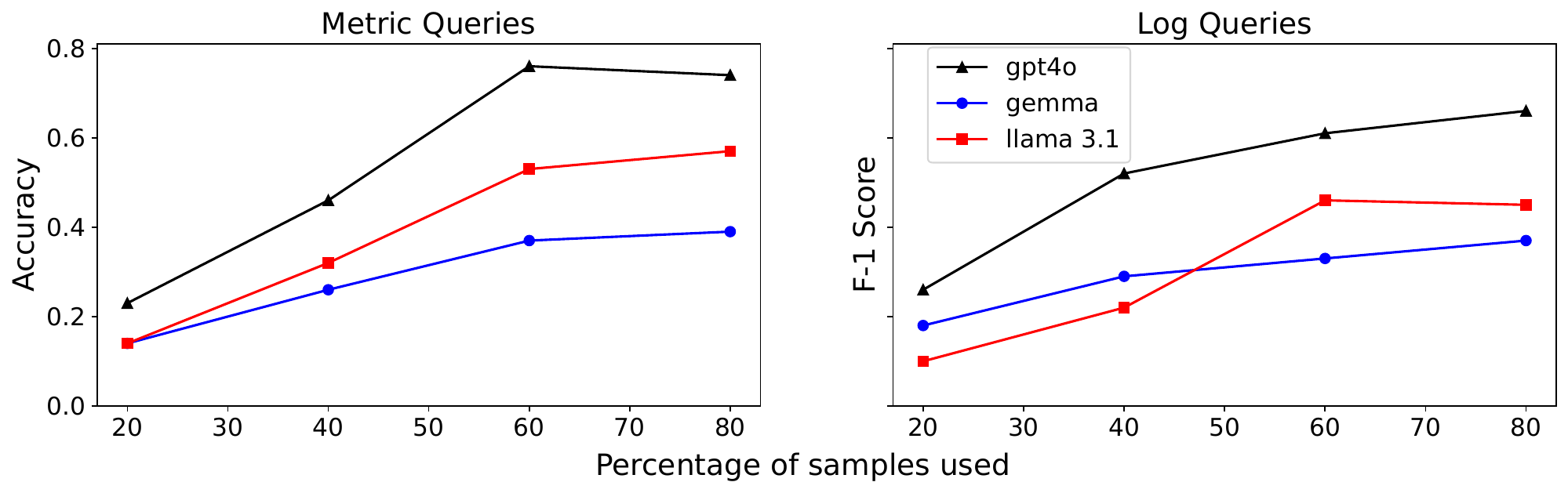}
\caption{OSTK}
\end{subfigure}
\begin{subfigure}{\linewidth}
\centering
\includegraphics[width=\textwidth]{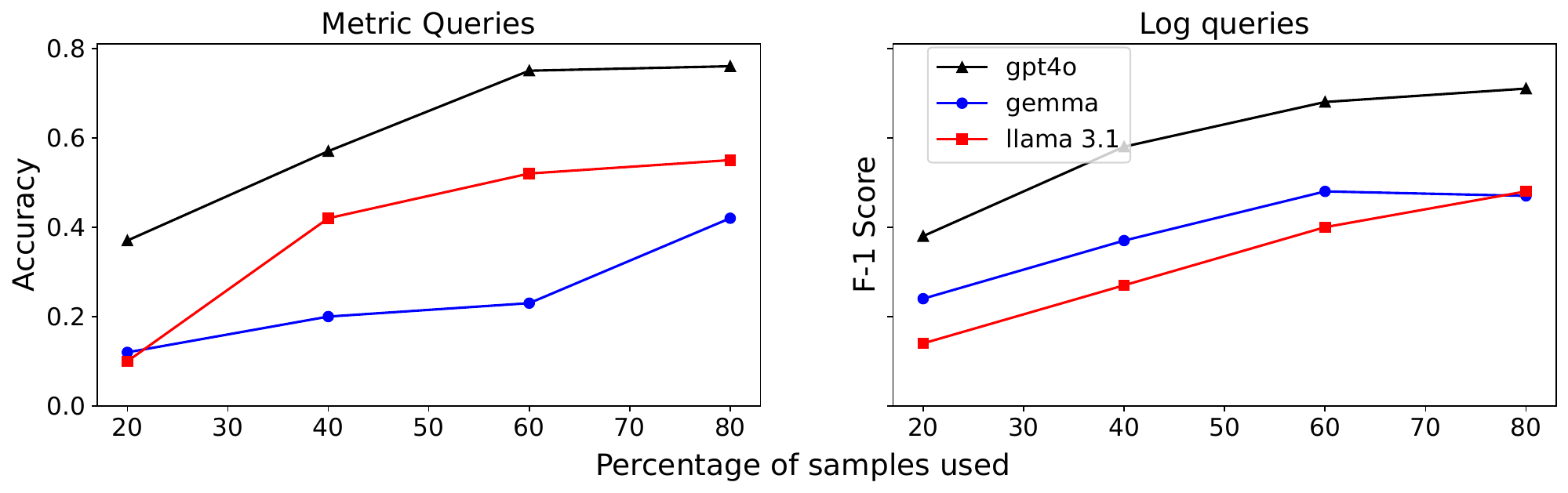}
\caption{OSSH}
\end{subfigure}
\begin{subfigure}{\linewidth}
\centering
\includegraphics[width=\textwidth]{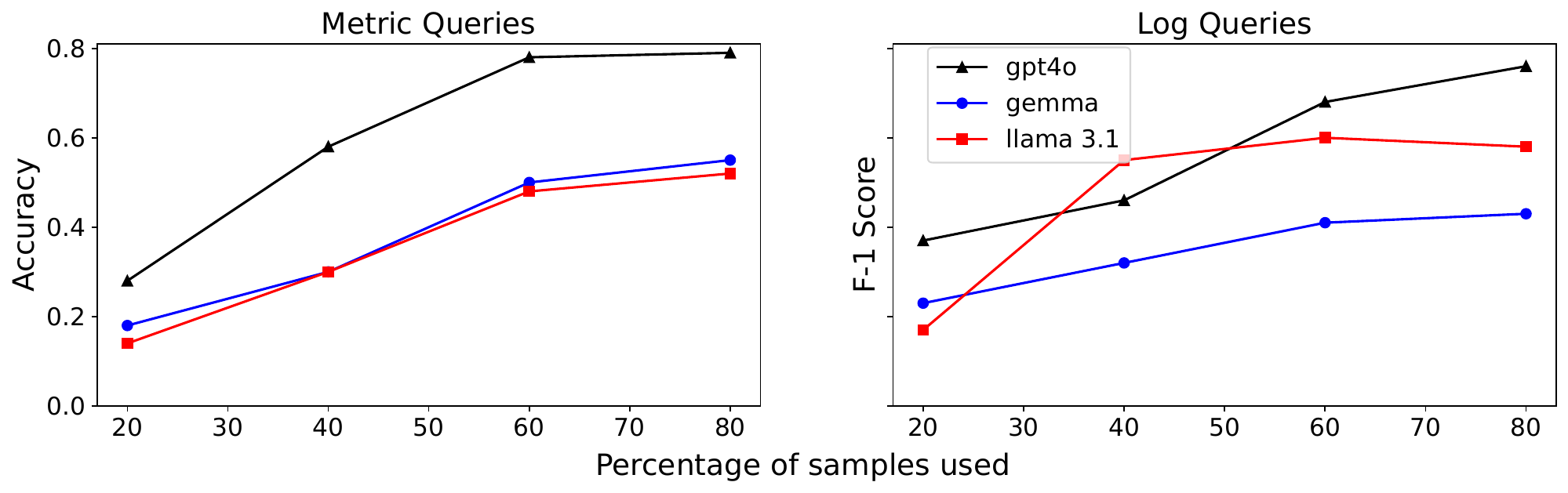}
\caption{HDFS}
\end{subfigure}
\caption{Model accuracy with different number of samples in finetuning phase
}
\label{fig:noofftsamples}
\end{figure*}

In the previous experiment, we looked into the effect of finetuning for enhancing the ability of the models to generate \logql queries compared to the base model. 
Previous works from other log related tasks \cite{llmparser2024ma}, and text2sql \cite{sqlauthoring2024Maddila} have shown that the performance of models change based on the number of samples used for finetuning. 
Constructing the dataset for finetuning these models is an arduous task as detailed in the previous section (\S\ref{sec:system_new:dataset}), thus we explore the effect of varying the number of samples used for finetuning the model. 
To perform this experiment, we allocated 20\% of samples from each application as a ``test set'', and remaining dataset for training the finetuned models. 
The number of finetuning samples were to be 20\%, 40\%, 60\% and 80\% of the overall sample, and the models were tested on the ``test set'' to obtain the metrics. 

\autoref{fig:noofftsamples} contains the analysis of model performance across varying finetuning sample sizes reveals a consistent pattern of improvement followed by plateau across all three models - Gemma, Llama 3.1, and GPT-4o. 
For metric queries, GPT-4o demonstrates superior performance, with accuracy increasing from 0.23-0.37 at 20\% samples to 0.74-0.79 at 80\% samples across applications. 
Similarly for log queries, GPT-4o achieves F1-scores ranging from 0.26-0.38 with 20\% samples, improving to 0.66-0.76 with 80\% samples. 
The performance gains are most pronounced when increasing from 20\% to 60\% of the training data, after which the improvements become marginal. 
For instance, GPT-4o's accuracy on HDFS metrics increases substantially from 0.28 (20\%) to 0.78 (60\%) but only marginally to 0.79 (80\%). 
This plateau effect is consistent across models and applications, suggesting that around 60\% of the training data captures most of the log formats and lines that are present in the logs for a particular application. 

Looking at individual models, Gemma shows the most modest improvements, with metric accuracy increasing from 0.12-0.18 at 20\% to 0.39-0.55 at 80\% across applications. 
Its F1-scores follow a similar trend, rising from 0.18-0.24 to 0.37-0.47. 
Llama 3.1 demonstrates slightly better scaling, achieving accuracies of 0.52-0.57 and F1-scores of 0.45-0.58 at 80\% samples, up from 0.10-0.14 and 0.10-0.17 respectively at 20\%. 
GPT-4o consistently outperforms both models across all sample sizes and applications. 
The HDFS application generally sees better performance compared to OSTK and OSSH across all models, particularly in the higher sample percentage ranges. 
The consistent plateauing behavior across all models and applications suggests an inherent limit to how much performance can be improved simply by increasing the amount of finetuning data. The difference in the results as more samples can be attributed to the LLMs understanding more information from the logs, which was a common problem in models pre-finetuning. 
Since each LLM learns the log patterns differently, there is a need for understanding the amount of samples required to finetune the LLMs, and future studies need to be conducted on understanding the interactions between the LLM, log query language and the amount of data required to finetune these models. 

\cemph{
Increasing the number of finetuning samples generally improves model performance in generating LogQL queries, with most gains achieved by 60\% of the training data, after which returns diminish significantly. While GPT-4o consistently outperforms Llama 3.1 and Gemma across all sample sizes, all models exhibit similar plateauing behavior, suggesting an inherent limit to performance improvements through increased training data alone.}

\subsection{Transferability of the finetuned models}
\label{sec:evaluation:transferability}

To evaluate the models' generalization capabilities across different applications, we conducted cross-application experiments where models were finetuned on two applications and subsequently tested on a third, previously unseen application. 
For instance, to assess query generation capabilities for OSSH logs, the models underwent finetuning using query datasets from OSTK and HDFS applications, thereby testing their ability to transfer learned patterns to a novel application context.

\begin{table}[h]
\begin{tabular}{|l|cc|cc|cc|}
\hline
\multirow{2}{*}{Model Name} & \multicolumn{2}{c|}{OSSH}& \multicolumn{2}{c|}{OSTK}  & \multicolumn{2}{c|}{HDFS} \\ \cline{2-7} 
& \multicolumn{1}{l|}{Metric} & \multicolumn{1}{l|}{Log} & \multicolumn{1}{l|}{Metric} & \multicolumn{1}{l|}{Log} & \multicolumn{1}{l|}{Metric} & \multicolumn{1}{l|}{Log} \\ \hline
Llama & \multicolumn{1}{r|}{0.13} & 0.14 & \multicolumn{1}{r|}{0.11}   & 0.07& \multicolumn{1}{r|}{0.05}   & 0.1\\ \hline
Gemma & \multicolumn{1}{r|}{0.09} & 0.07 & \multicolumn{1}{r|}{0.12}& 0.18 & \multicolumn{1}{r|}{0.09} & 0.16 \\ \hline
GPT-4o & \multicolumn{1}{r|}{\textbf{0.32} } & \textbf{0.33} & \multicolumn{1}{r|}{\textbf{0.22}} & \textbf{0.47} & \multicolumn{1}{r|}{\textbf{0.27}} &\textbf{0.29} \\ \hline
\end{tabular}
\caption{Results for transferability of finetuned models across applications.  
}
\label{tab:finetuning_transferability}
\end{table}

Table \ref{tab:finetuning_transferability} presents the performance metrics of models after finetuning on two applications and evaluating on a third application. The results reveal that while cross-application fine-tuned models are inferior to application-specific fine-tuned models in most cases, they generally outperform their non-finetuned counterparts.

For example, GPT-4o demonstrates strong performance across all applications, achieving metric accuracy between 0.22 and 0.32 and log query F1-scores ranging from 0.29 to 0.47. Importantly, fine-tuning GPT-4o on OSTK and HDFC leads to significant relative improvements as compared to the non-fine-tuned counterparts, with performance gains of 52\% (from 0.21 to 0.31) and 106\% (from 0.16 to 0.33) in OSSH metric accuracy and log queries, respectively.

Moreover, smaller models like Llama and Gemma show minor improvements with respect to their non-fine-tuned versions. Llama's metric accuracy ranges from 0.13 on OSSH to 0.05 on HDFS, while Gemma shows inconsistent metric accuracy, ranging between 0.09 and 0.12. These marginal improvements in accuracy and F1-scores can be attributed to the models’ ability to capture syntactic patterns from the fine-tuning dataset. 

However, the overall limited performance of these models stems from insufficient exposure to application-specific log patterns and their corresponding log query labels during training. Consequently, the errors observed in these models mirror those of their non-finetuned versions, particularly in their inability to effectively incorporate application-specific log information.

\cemph{Evaluation of cross-application transferability revealed that models finetuned on two applications demonstrate limited performance when tested on a third, unseen application. Despite GPT-4o showing relatively better performance with metric accuracy up to 0.32, all models exhibited performance levels closer to their non-finetuned versions, primarily due to insufficient exposure to application-specific log patterns and corresponding query labels.}

\subsection{Code Quality Analysis}

\begin{table}[]
\begin{tabular}{|l|r|r|r|}
\hline
Model & OSSH & OSTK & HDFS \\ \hline
Llama      & 0.57                     & 0.58                     & 0.59                     \\
Gemma      & 0.43                     & 0.39                     & 0.46                     \\
GPT-4o     & \textbf{0.78}                    & \textbf{0.86}                     & \textbf{0.77}                     \\ \hline
\end{tabular}
\caption{Codebert score for various models and applications 
}
\label{tab:codebertscore}
\end{table}

For analyzing the quality of the code produced by various finetuned models, we make use of CodeBertScore\cite{codebertscore2023Zhou} (CBS) to evaluate the quality of the \logql query generated by the finetuned models. 

\textbf{Finetuning CBS} Since the current CBS model doesn't support \logql, we finetuned the CBS model to be able to score the outputs for the model. 
We used 50\% of application specific \logql queries to finetune the CBS model, and used 30\% of the dataset to finetune the LLMs, and tested it on 20\% of the dataset by comparing the output of the finetuned models with the correctly written \logql queries in the dataset. 
Table \ref{tab:codebertscore} shows the CBS for the queries generated by the finetuned models. 
GPT-4o  consistently achieves the highest CodeBERTScore across all applications, with scores of 0.78 for OpenSSH (OSSH), 0.86 for OpenStack (OSTK), and 0.77 for HDFS. 
These high scores indicate that GPT-4o generates LogQL queries that are highly similar to the reference queries, both semantically and functionally. 
In contrast, the Gemma model exhibits significantly lower scores, ranging from 0.39 to 0.46, suggesting that its generated queries are less aligned with the reference queries and may require substantial refinement to achieve functional correctness. 
The Llama model performs moderately, with scores between 0.57 and 0.59, indicating partial similarity but room for improvement in terms of query accuracy and efficiency.
Overall, these results highlight the importance of selecting models with higher CodeBERTScores to ensure the generation of high-quality LogQL queries that are both syntactically correct and functionally reliable. 

\cemph{
The evaluation demonstrates that finetuned LLM models can successfully generate LogQL queries, with varying degrees of accuracy across different models.
GPT-4o emerged as the top performer, whereas other models showed moderate to lower performance, with Llama achieving scores between 0.57-0.59 and Gemma scoring between 0.39-0.46, suggesting their generated queries require more refinement to achieve full functional reliability.
}

\section{Discussion}
\label{sec:discussion}

This work showed that there is a necessity for enhanced interfaces in observability data query generation. 
To our knowledge, this research presents the first comprehensive data collection effort for fine-tuning models to generate log query language. 
Our evaluations demonstrate that while base LLMs exhibit limitations in generating \logql queries, fine-tuning these models significantly enhances their query generation capabilities. 
Although we present a proof of concept across various applications, practitioners seeking to implement our methodology would need to develop a corpus of natural language to \logql queries specific to their applications, as accurate query generation necessitates understanding application-specific log semantics. 
Additionally, the model must generate syntactically valid queries compatible with their internal system's query language. 
These considerations directly influence the selection of base models for fine-tuning purposes. 
For instance, organizations utilizing Datadog for observability data storage would need to develop a dataset mapping Natural Language to DQL queries, and select base models capable of DQL query generation. 
This dataset must encompass diverse query types addressing the three domains outlined in Section \ref{sec:system_new:dataset}. 
This research aims to serve as a framework for organizations developing observability data queries within their specific contexts.

\subsection{Threats to Validity}
The fine-tuned \sysName{} models demonstrate superior performance metrics across both metric and log queries compared to the baseline model. 
However, as this research represents one of the initial endeavors in \logql query generation, several limitations, and opportunities for improvement warrant discussion. 

A significant limitation observed in the current implementation is the models' reduced efficacy in generating \logql queries when there exists semantic divergence between the natural language (NL) query and the corresponding log entries. 
To illustrate, consider the example in Table \ref{tab:example_dataset}-row 3, where the objective is to identify recent successful authentication events for user ``fztu''. 
While the NL query employs the phrase ``successful login,'' the actual log entries utilize ``accepted password'' to denote such events. 
Consequently, the model frequently generates queries containing ``successful login,'' resulting in null result sets.
This limitation can be attributed to two primary factors. 
First, the methodology of reverse-engineering natural language (NL) questions from existing dashboards inherently limits the diversity of NL queries that can map to a specific \logql query. Contemporary Text-to-SQL benchmarks attempt to address this limitation by manually writing multiple natural language questions for a SQL query, which can easily get cumbersome. 
One potential ancillary direction is to design automated paraphrasing techniques to generate semantically equivalent and useful NL queries. 
Second, current log aggregation systems, including Loki, lack semantic search capabilities within log data. 
The implementation of semantic similarity matching in search systems would enable \logql queries containing phrases like ``successful login'' to successfully retrieve log entries containing semantically equivalent terms such as ``accepted password.''

An interesting direction for future research is exploring how to achieve additional performance gains in generating viable \logql queries for application logs absent from the fine-tuning dataset. As demonstrated in Section \ref{sec:evaluation:transferability}, while fine-tuning successfully addresses certain syntactical challenges, enhancing model transferability necessitates the incorporation of diverse application logs during the fine-tuning process. 
This diversity spans across logging frameworks, application domains, and implementation patterns, creating substantial complexity in sampling representative log entries. 
Although utilizing application logs for fine-tuning could reduce dependence on manually created NL-to-\logql pairs, the challenge of sampling a sufficiently diverse and representative set of logs remains an open research problem~\cite{llmparser2024ma}.

Our work focuses on generating \logql queries directly from natural language, presenting an alternative to the widely used practice of constructing \logql queries through Grafana dashboards. The existing workflow, which requires users to migrate queries between our interface and Grafana, introduces significant cognitive overhead due to constant context switching and hampers developer productivity. To address this, we envision an intelligent interface, akin to GitHub Copilot, that provides contextual suggestions for query entities, labels, and log lines during composition. However, realizing this vision involves several technical challenges: (a) the need for extensive training data mapping natural language to \logql queries, (b) meeting stringent latency requirements for real-time autocompletion, and (c) supporting bidirectional context processing, which diverges from the conventional left-to-right generation approach in NL-to-\logql systems. Overcoming these challenges would require restructuring our dataset to focus on \logql queries and adopting a Fill in the Middle (FIM) architecture to facilitate the predictive completion of query prefixes and suffixes during interactive development.

\section{Conclusion}
\label{sec:conclusion}
This research addresses the Natural Language to LogQL (NL2LogQL) problem, focusing on the generation of log-file queries from natural language input. 
Our approach advances the state of the art by eliminating two significant prerequisites: the requirement for users to possess expertise in \logql query language, and the necessity for detailed understanding of log file structure and syntax. 
To facilitate this advancement, we developed a comprehensive dataset comprising 424 natural language queries, their corresponding \logql implementations, and their expected query results across extensive log files. 
Our methodology leverages fine-tuned large language models, selected for their dual capabilities: (1) inherent comprehension of programming languages, query syntax, and \logql, enabling the generation of syntactically and semantically valid queries, and (2) efficient processing and interpretation of extensive log files for understanding structural patterns and definitions. 
Empirical evaluation of our \sysName system demonstrates substantial improvements in performance post-fine-tuning, achieving a success rate exceeding 75\% in generating syntactically and semantically correct \logql queries. 
These generated queries produce results comparable to those crafted by human experts possessing dual expertise in both \logql query language and log file interpretation, validating the effectiveness of our approach.

\bibliographystyle{ACM-Reference-Format}
\bibliography{vish_llm_project}

\end{document}